\documentstyle[12pt]{article}
\title{Separation of variables in the Kramers equation}
\author{Renat Zhdanov\thanks{e-mail: renat@imath.kiev.ua}\ \
and Alexander Zhalij\thanks{e-mail: zhaliy@imath.kiev.ua} \\ \small Institute
of Mathematics of the Academy of Sciences of Ukraine,\\
\small Tereshchenkivska Street 3, 252004 Kyiv, Ukraine}
\date{}
\newtheorem{theo}{Theorem}
\let\dis\displaystyle
\begin{document}
\maketitle
\begin{abstract}
  We consider the problem of separation of variables in the Kramers
  equation admitting a non-trivial symmetry group. Provided the
  external potential $V(x)$ is at most quadratic, a complete solution
  of the problem of separation of variables is obtained. Furthermore,
  we construct solutions of the Kramers equation with separated
  variables in explicit form.
\end{abstract}
\section{Introduction}

Many phenomena in physics and, especially, in chemical physics may
be modelled as the Brownian motion of particles in an external
potential $V(x)$, the appropriate transport equation being the
(1+2)-dimensional Fokker-Plank equation of special form
\begin{equation}
\label{0.0}
u_t=\nu u_{yy} - yu_x +(\nu y + V'(x))u_y + \nu u,
\end{equation}
where $u=u(t,x,y)$ is a sufficiently smooth real-valued function and
$\nu$ is a real parameter.

The first relevant result on studying partial differential equation
(PDE) (\ref{0.0}) has been obtained by Kramers \cite{kra40}. He found a
solution of the escape problem of a classical particle subjected to
Gaussian white noise out of a deep potential well. This is why the
equation in question is called the Kramers equation (KE) (see, for more
details \cite{gar85}--\cite{fer93}).

As KE is a PDE with variable coefficients, we
cannot apply the Fourier transform in order to solve it. In fact the
only way to obtain exact solutions of KE are either to utilize
its Lie symmetry or to apply the method of separation of variables. The
first possibility has been exploited recently in \cite{spi97,spi98},
where symmetry classification of the class of PDEs (\ref{0.0}) has been
carried out. The principal result of these  papers is that KE has a
symmetry group that is wider than a trivial one-parameter group of time
translations if and only if $V''(x)=0$.

The principal aim of the present paper is to apply the direct approach
to variable separation in PDEs suggested in
\cite{zhd93}--\cite{zhd97} to solve KE. As is well-known, separability
of PDE is  intimately connected to its symmetry within the class of
second-order differential operators \cite{mil77}. This is why, we will
concentrate on the case $V(x)=kx,\ k={\rm constant}$, namely, we consider
KE having non-trivial Lie symmetry
\begin{equation}
\label{0.1}
u_t=\nu u_{yy} - yu_x +(\nu y + kx)u_y + \nu u.
\end{equation}

In a classical setting the method of separation of variables
(say, in the Cartesian coordinate system) is based on a special
representation of a solution to be found in factorized form:
\[
u(t,x,y)=\varphi_0(t)\varphi_1(x)\varphi_2(y),
\]
where $\varphi_i, i=0,1,2$ are solutions of some ordinary differential
equations (ODEs). However, one can try to separate variables in the equation
under study in another coordinate system, for example, in
polar coordinates and look for a solution of the form
\[
u(t,x,y)=\varphi_0(t)\varphi_1\left(\sqrt{x^2+y^2}\right)
\varphi_2\left(\arctan {y\over x}\right).
\]
So, if we are given some coordinate system, then it is clear how
to get exact solutions with separated variables. However, the
classical approach gives no general routine for finding all possible
coordinate systems providing separability of the equation under study.
Our approach to the problem of the separation of variables in evolution-type
equations (to be specific, we take the case of an equation
having three independent variables $t, x, y$) is based on the following
observations:
\begin{itemize}
\item{All solutions with separated variables known to us can be
represented in the form
\begin{equation}
\label{0.2}
u(t,x,y)=Q(t,x,y)\varphi_0(t)\varphi_1(\omega_1(t,x,y))
\varphi_2(\omega_2(t,x,y)),
\end{equation}
where $Q, \omega_1, \omega_2$ are sufficiently smooth functions and
$\varphi_i, i=0,1,2$ satisfy some ODEs.}
\item{The functions $\varphi_i, i=0,1,2$ depend on two arbitrary
parameters $\lambda_1, \lambda_2$ called spectral parameters or separation
constants. Furthermore, the functions $Q, \omega_1, \omega_2$ are
independent of $\lambda_1, \lambda_2$.}
\end{itemize}

By properly postulating these features we have formulated an
efficient approach to the problem of variable separation in linear
PDEs \cite{zhd97}. Applying it to the KE (\ref{0.1})
we look for its particular solutions of the
form (\ref{0.2}), where functions $Q, \omega_1, \omega_2$ are chosen in such
a way that inserting (\ref{0.2}) into KE yields three ODEs for
functions $\varphi_0(t), \varphi_1(\omega_1), \varphi_2(\omega_2)$
\begin{equation}
\label{0.3}
\begin{array}{l}
U_0(t, \varphi_0, \dot \varphi_0; \lambda_1, \lambda_2)=0,\\[2mm]
U_i(\omega_i, \varphi_i, \dot \varphi_i, \ddot \varphi_i; \lambda_1,
\lambda_2)=0,\quad
i=1,2.
\end{array}
\end{equation}
Here $U_0, U_1, U_2$ are some smooth functions of the indicated
variables, $\lambda_1, \lambda_2$ are real parameters and, what is more,
\begin{equation}
\label{rank}
{\rm rank}\, \left \|\begin{array}{cc}
{\partial U_0\over \partial \lambda_1}&{\partial U_0\over \partial \lambda_2}\\[3mm]
{\partial U_1\over \partial \lambda_1}&{\partial U_1\over \partial \lambda_2}\\[3mm]
{\partial U_2\over \partial \lambda_1}&{\partial U_2\over \partial \lambda_2}
\end{array}\right\|=2.
\end{equation}
Note that the functions $Q, \omega_1, \omega_2$ are independent
of $\lambda_1, \lambda_2$.

Provided the above requirements are met, we say that KE is separable in
the coordinate system $t, \omega_1(t,x,y), \omega_2(t,x,y)$.

Due to the fact that the equation under study is linear, the reduced
equations prove to be linear as well. Furthermore, we have to consider
two distinct cases.
\vspace{2mm}

\noindent
{Case 1.} The system of equations (\ref{0.3}) has the form
\begin{eqnarray}
\dot\varphi_0&=&A_0(t; \lambda_1, \lambda_2)\varphi_0,\nonumber\\
\dot\varphi_1&=&A_1(\omega_1; \lambda_1, \lambda_2)\varphi_1,\label{0.4}\\
\ddot\varphi_2&=&A_2(\omega_2; \lambda_1, \lambda_2)\dot\varphi_2+
                 A_3(\omega_2; \lambda_1, \lambda_2)\varphi_2. \nonumber
\end{eqnarray}
{Case 2.} The system of equations (\ref{0.3}) has the form
\begin{eqnarray}
\dot\varphi_0&=&A_0(t; \lambda_1, \lambda_2)\varphi_0,\nonumber\\
\dot\varphi_1&=&A_1(\omega_1; \lambda_1, \lambda_2)\varphi_1,\label{0.5}\\
\dot\varphi_2&=&A_2(\omega_2; \lambda_1, \lambda_2)\varphi_2. \nonumber
\end{eqnarray}
In these formulae $A_0,\ldots, A_3$ are some smooth real-valued
functions of the indicated variables.

Consequently, there are two different possibilities to separate
variables in KE, either to reduce it to two first-order and one
second-order ODEs or to three first-order ODEs. It is impossible to
reduce KE to two or three second-order ODEs because it contains a
second-order derivative with respect to one variable only.

Provided the system of reduced ODEs has the form (\ref{0.4}), separation
of variables in (\ref{0.1}) is performed in the following way:
\begin{enumerate}
\item{We insert the Ansatz (\ref{0.2}) into KE and express the
derivatives $\dot \varphi_0$, $\dot \varphi_1$, $\ddot \varphi_1$,
$\ddot \varphi_2$ in terms of functions $\varphi_0, \varphi_1,
\varphi_2, \dot \varphi_2$ using equations (\ref{0.4}) and their
differential consequences (where necessary).}
\item{The equality obtained is split by $\varphi_0, \varphi_1,
\varphi_2, \dot \varphi_2, \lambda_1, \lambda_2$ which are regarded as
independent variables. This yields an over-determined system of
nonlinear PDEs for unknown functions $Q, \omega_1, \omega_2$.}
\item{After solving the above system we get an exhaustive description
of coordinate systems providing separability of KE.}
\end{enumerate}

Clearly, if we adopt a more general definition of the separation of
variables, then additional coordinate systems providing separability of
KE may appear. However, all solutions with separated variables of the
Schr\"odinger and heat conductivity equations known to us can be
obtained within the described approach.

The case when the system of reduced ODEs is of the form (\ref{0.5}) is
handled in a similar way.

Next, we introduce an equivalence relation on the set of all coordinate
systems providing separability of KE. We say that two coordinate systems
$t,\ \omega_1,\ \omega_2$ and $t',\ \omega'_1,\ \omega'_2$ are
equivalent if the corresponding solutions with separated variables are
transformed  one into another by
\begin{itemize}
\item{the group transformations from the Lie transformation group
admitted by KE,}
\item{the transformations of the form
\begin{eqnarray}
&&t\to t'=f_0(t),\quad \omega_i\to \omega'_i=f_i(\omega_i),\label{0.6a}\\
&&Q\to Q'=Qh_0(t) h_1(\omega_1) h_2(\omega_2),\label{0.6b}
\end{eqnarray}
where $f_0, f_i, h_0, h_i$ are some smooth functions.}
\end{itemize}

It can be proved that formulae (\ref{0.6a}), (\ref{0.6b}) define the
most general transformation preserving the class of Ans\"atze (\ref{0.2}).
The equivalence relation split the set of all possible coordinate
systems into equivalence classes. In a sequel, when presenting the lists
of coordinate systems enabling us to separate variables in KE we will
give only one representative for each equivalence class.

\section{Principal results}
In this section we give a complete account of our results on
the separation of variables in KE obtained within the framework of the
approach described in Introduction. We write down explicit forms of
the functions $Q(t,x,y)$, $\omega_1(t,x,y)$, $\omega_2(t,x,y)$ and the
corresponding reduced ODEs for functions $\varphi_0(t)$,
$\varphi_1(\omega_1)$, $\varphi_2(\omega_2)$.
\begin{theo}
Equation (\ref{0.1}) admits the separation of variables into two first-order and
one second-order ODEs if and only if $k$ takes one of the three values
$0,\ 3\nu^2/16,\ -3\nu^2/4$. Furthermore, equations separates into three
first-order ODEs with an arbitrary $k$.
\end{theo}

Theorem 1 gives a general description of separable KEs. The
solution of the problem of separation of variables in corresponding KEs is
pro\-vi\-ded by Theorems 2--6 later.
\begin{theo}
The set of inequivalent coordinate systems providing separability of KE with $k=\nu^2/4$
is exhausted by the following ones:
\begin{equation}\label{1.2}
\begin{array}{l}
\omega_i={\dis f_iy-\dot f_ix\over\dis \dot f_2f_1-\dot f_1f_2},\quad i=1,2,\\[5mm]
Q=\exp\Biggl\{ \left(\dis
{-{1\over 4\nu}{\ddot f_2f_1-\ddot f_1f_2\over \dot f_2
f_1-\dot f_1f_2}-{1\over 4}}\right)y^2+{\dis {1\over 2\nu}} \left({\dis
{\ddot f_2\dot f_1-\ddot f_1\dot f_2\over \dot f_2f_1-\dot f_1f_2}-k}\right)
x\\[6mm]
\quad +\left(\dis {{1\over 4\nu}{\stackrel{...}{f_2}\dot f_1-\stackrel{...}{f_1}\dot
f_2\over \dot f_2f_1-\dot f_1f_2}-{k\over 4}}\right)x^2-{\dis{1\over 2}}
\ln|
 \dot f_2f_1-\dot f_1f_2|+{\dis{\nu\over 2}}t\Biggl\};\\[5mm]
\dot\varphi_0 =\nu\left(\dis {f_1\lambda_1+f_2\lambda_2\over  \dot f_2f_1-\dot f_1
f_2}\right)^2\varphi_0,\quad \dot \varphi_1 =\lambda_1\varphi_1,\quad
\dot \varphi_2 =\lambda_2\varphi_2,
\end{array}
\end{equation}
where
\begin{eqnarray*}
f_1&=&t\left(A_1\sinh{\nu\over 2}t+A_2\cosh{\nu\over 2}t\right)+
A_3\sinh{\nu\over 2}t+A_4\cosh{\nu\over 2}t,\\
f_2&=&t\left(B_1\sinh{\nu\over 2}t+B_2\cosh{\nu\over 2}t\right)+
B_3\sinh{\nu\over 2}t+B_4\cosh{\nu\over 2}t,
\end{eqnarray*}
and $A_1,\ldots, B_4$ are arbitrary real constants satisfying the
condition $2C_{12}$ $-\nu (C_{13}$ $-C_{24})$$=0$.
 Hereafter we use the notations
\[
C_{ij}=B_iA_j - A_iB_j,\quad i,j=1,\ldots, 4.
\]
\end{theo}

\begin{theo}
The set of inequivalent coordinate systems providing separability of KE with $k>\nu^2/4$
is exhausted by those given in (\ref{1.2}) with
$$f_1 =\sin bt(A_1\sinh at+A_2\cosh at)+\cos bt(A_3\sinh at+A_4\cosh at),$$
$$f_2 =\sin bt(B_1\sinh at+B_2\cosh at)+\cos bt(B_3\sinh at+B_4\cosh at),$$
where $a={\nu\over 2},b=\sqrt{k-{\nu^2\over 4}}$ and $A_1,\ldots, B_4$
are constants fulfilling the condition $(C_{12}+C_{34})b+(C_{13}-C_{24})a=0$.
The explicit form of the function $Q$ and the reduced ODEs are also obtained
from the formulae (\ref{1.2}) with $f_1, f_2$ given previously.
\end{theo}
\begin{theo}
The set of inequivalent coordinate systems providing separability of KE
with $k<\nu^2/4$ and $k\ne 0,\ 3\nu^2/16,\ -3\nu^2/4$ is exhausted by those
given in (\ref{1.2}) with
$$f_1=\sinh bt(A_1\sinh at+A_2\cosh at)+\cosh bt(A_3\sinh at+A_4\cosh at),$$
$$f_2=\sinh bt(B_1\sinh at+B_2\cosh at)+\cosh bt(B_3\sinh at+B_4\cosh at),$$
where $a={\nu\over 2},b=\sqrt{{\nu^2\over 4}-k}$ and $A_1,\ldots, B_4$
are constants fulfilling the condition $(C_{12}-C_{34})b+(C_{13}-C_{24})a=0$.
The explicit form of the function $Q$ and reduced ODEs are also obtained
from the formulae (\ref{1.2}) with $f_1, f_2$ given before.
\end{theo}

\begin{theo}
The set of inequivalent coordinate systems providing separability of KE with $k=0$
is exhausted by
\begin{enumerate}
\item[{1)}]{those given in (\ref{1.2}) with
$$f_1=A_1\sinh\nu t+A_2\cosh\nu t+A_3t+A_4,$$
$$f_2=B_1\sinh\nu t+B_2\cosh\nu t+B_3t+B_4,$$
where $A_1,\ldots, B_4$ are constants fulfilling the equation
$\nu C_{12} - C_{34} = 0$.
The explicit form of the function $Q$ and reduced ODEs are also obtained
from the formulae (\ref{1.2}) with $f_1, f_2$ given previously;}
\item[{2)}]{the following coordinate system:
\begin{eqnarray*}
&&\omega_1 =x,\quad \omega_2=y,\quad Q=\exp\left(\dis -{y^2\over 4}\right);\\
&&\dot \varphi_0 =\nu\lambda_1\varphi_0,\quad
\dot \varphi_1 =\nu\lambda_2\varphi_1,\quad
\ddot \varphi_2 =\left({y^2\over 4}+\lambda_2y+\lambda_1-{1\over 2}\right)\varphi_2.
\end{eqnarray*}}
\end{enumerate}
\end{theo}

\begin{theo}
The set of inequivalent coordinate systems providing separability of KE
with $k=3\nu^2/16$ or $k=-3\nu^2/4$ is exhausted by
\begin{enumerate}
\item[1)]{those given in Theorem 4 under $k=3\nu^2/16$ or $k=-3\nu^2/4$;}
\item[2)] {the following coordinate systems:
\begin{eqnarray*}
&&\omega_1=R^3x,\quad \omega_2=Ry+3\dot Rx,\\
&&Q=\exp\Biggl\{\left(\dis {{\dot R\over \nu R}-{1\over 4}}\right)y^2+{1\over 2\nu}
\left({3{\ddot R\over R}-k}\right)xy+\left(-{3\stackrel{...}{R}\over 4\nu R}+
{15\dot R\ddot R\over 4\nu R^2}\right.\\
&& \quad \left.-{k\over 4}\right)x^2+{\nu\over 2}t+2\ln R\Biggr\};\\
&&\dot\varphi_0=\nu\lambda_1R^2\varphi_0,\quad
\dot\varphi_1=\nu\lambda_2\varphi_1,\quad
\ddot\varphi_2=(\lambda_2\omega_2+\lambda_1)\varphi_2,
\end{eqnarray*}
where
\[
R(t)=\left \{\begin{array}{l} \frac{\dis 1}{\dis\cosh at},\\[3mm]
                              \frac{\dis 1}{\dis\sinh at},\\[3mm]
                              \exp \{\pm at\} \end{array}\right.\quad {\rm with}\quad
a=\left \{\begin{array}{ll} {\dis \frac{\nu}{4}},& {\it under}\
                            k={\dis \frac{3\nu^2}{16}},\\[3mm]
                            {\dis \frac{\nu}{2}},& {\it under}\
                            k=-{\dis \frac{3\nu^2}{4}}.
                            \end{array}\right.
                            \]}
\end{enumerate}
\end{theo}

\section{Proof of Theorems 1--6}

In order to prove the assertions of the previous section one should apply
to equation (\ref{0.1}) the algorithm of variable separation described
in Introduction.

We give a detailed proof for the case when system of reduced ODEs is
of the form (\ref{0.4}). Inserting Ansatz (\ref{0.2}) into KE
(\ref{0.1}) and expressing the derivatives $\dot \varphi_0$, $\dot
\varphi_1$, $\ddot \varphi_1$, $\ddot \varphi_2$ in terms of functions
$\varphi_0, \varphi_1, \varphi_2, \dot \varphi_2$ with the use of
equations (\ref{0.4}) and their differential consequences yield a
system of two nonlinear PDEs
\begin{eqnarray}
&&Q\omega_{2t}+yQ\omega_{2x}=\nu(yQ\omega_{2y}+2Q_y\omega_{2y}+2QA_1\omega_{1y}
\omega_{2y}\nonumber\\
&&\quad +QA_2\omega^2_{2y}+Q\omega_{2yy})+kxQ\omega_{2y} , \label{2.1}\\
&&Q_t+QA_0+QA_1\omega_{1t}+yQ_x+yQA_1\omega_{1x}=\nu(Q+yQ_y\nonumber\\
&&\quad +yQA_1\omega_{1y}+Q_{yy}+2Q_yA_1\omega_{1y}+Q(A^2_1+A_{1\omega_1})
\omega^2_{1y}\nonumber\\
&&\quad +QA_1\omega_{1yy}+QA_3\omega^2_{2y})+kx(Q_y+QA_1\omega_{1y}).  \label{2.2}
\end{eqnarray}

This system is to be split with respect to variables $\lambda_1,
\lambda_2$ (we remind that the functions $\omega_1, \omega_2$ are
independent of $\lambda_1, \lambda_2$). To this end we differentiate
(\ref{2.1}) with respect to $\lambda_i$ and get the relation
\[
(2A_{1\lambda_i}\omega_{1y} + A_{2\lambda_i}\omega_{2y})\omega_{2y}=0,\quad
i=1,2
\]
Due to the fact that $\omega_{2y}$ does not vanish identically (otherwise
it follows from (\ref{2.1}) that $\omega_2=$constant), the equation
\begin{equation}
\label{2.3}
2A_{1\lambda_i}\omega_{1y} + A_{2\lambda_i}\omega_{2y}=0,\quad i=1,2.
\end{equation}
holds.

Let us show first that we can, without loss of generality, put
$\omega_{1y}=0$. Suppose the inverse, namely that the inequality
$\omega_{1y}\ne 0$ holds true. It follows from the second equation
of system (\ref{0.4}) that $A_{1\lambda_1}^2+A_{1\lambda_2}^2\ne 0$.
Let the function $A_{1\lambda_1}$ be non-vanishing, then by the influence of
(\ref{2.3}) $A_{2\lambda_1}\ne 0$. Denoting
\[
A_{1\lambda_1}=g(\omega_1,\lambda_1,\lambda_2),\quad
-2A_{2\lambda_1}=f(\omega_2,\lambda_1,\lambda_2)
\]
we rewrite (\ref{2.3}) as follows
\begin{equation}
\label{2.4}
{\omega_{1y}\over \omega_{2y}}={f(\omega_2,\lambda_1,\lambda_2)\over
g(\omega_1,\lambda_1,\lambda_2)}.
\end{equation}

Differentiating (\ref{2.4}) with respect to $\lambda_1$ yields
\[
{f_{\lambda_1}\over f}= {g_{\lambda_1}\over g}.
\]
Hence we conclude that there is a function $k=k(\lambda_1,\lambda_2)$ such that
\[
{f_{\lambda_1}\over f}= {g_{\lambda_1}\over g}=k(\lambda_1,\lambda_2).
\]
Integrating the above equations we get
\[
f=k_1(\lambda_1,\lambda_2)f_1(\omega_2,\lambda_2),\quad
g=k_1(\lambda_1,\lambda_2)g_1(\omega_1,\lambda_1),
\]
so that (\ref{2.4}) reduces to the relation
\[
{\omega_{1y}\over \omega_{2y}}={f_1(\omega_2,\lambda_2)\over
g_1(\omega_1,\lambda_2)}.
\]
In a similar way we establish that the last relation is equivalent
to the following one
\[
{\omega_{1y}\over \omega_{2y}}={f_2(\omega_2)\over
g_2(\omega_1)},
\]
hence
\[
g_2(\omega_1) \omega_{1y} = f_2(\omega_2) \omega_{2y}.
\]

Taking into account the equivalence relation (\ref{0.6a}) we can
put $g_2=1$ and $f_2=1$ in the above equality thus getting $\omega_{1y}=
\omega_{2y}$. Integrating this PDE yields
\[
\label{2.5}
\omega_1=\omega_2 + h(t,x)
\]
with an arbitrary smooth function $h$. In view of this equation
relation (\ref{2.3}) takes the form
\[
2A_{1\lambda_i} + A_{2\lambda_i}=0,\quad i=1,2.
\]
Hence we conclude that there exists a function $\Lambda(\lambda_1, \lambda_2)$
such that
\begin{equation}
\label{2.6}
A_1=\Lambda(\lambda_1, \lambda_2) + \tilde A_1(\omega_1),\quad
A_2=-2\Lambda(\lambda_1, \lambda_2) + \tilde A_2(\omega_2).
\end{equation}
Within the equivalence transformation (\ref{0.6b}) with properly chosen functions
$h_1, h_2$ we can put $\tilde A_1(\omega_1)=0, \tilde A_2(\omega_2)=0$. Furthermore,
defining new separation constants as
\[
\lambda_1'=\Lambda(\lambda_1,\lambda_2),\quad \lambda_2'=\lambda_2
\]
and omitting the primes we represent (\ref{2.6}) in the form
\[
A_1=\lambda_1,\quad A_2=-2\lambda_1.
\]

Consequently, system (\ref{0.4}) takes the form
\begin{eqnarray}
\dot \varphi_0(t)&=& A_0(t)\varphi_0(t),\nonumber\\
\dot \varphi_1(\omega_1)&=&\lambda_1\varphi_1(\omega_1),\label{2.7}\\
\ddot\varphi_2(\omega_2)&=& -2\lambda_1\dot\varphi_2(\omega_2) +
A_3(\omega_2,\lambda_1,\lambda_2)\varphi_2(\omega_2).\nonumber
\end{eqnarray}

Making the change of variables $\varphi_2=\phi\exp\{-\lambda_1\omega_2\}$
reduces the third equation of system (\ref{2.7}) to
\[
\ddot \phi = (\lambda_1^2+A_3)\phi.
\]
Let $\phi=\phi(\omega_2,\lambda_1,\lambda_2)$ be a solution of this
equation. Then the corresponding solution with separated variables
becomes
\[
u=Q(t,x,y)\phi(\omega_2,\lambda_1,\lambda_2)\exp\left\{\int A_0(t){\rm d}t +
\lambda_1(\omega_1-\omega_2)\right\}.
\]
The structure of so obtained solution with separated variables is
such that dependence of $\omega_1$ on $y$ is inessential. Indeed,
the function $\omega_1$ enters into the solution only as a
combination $\omega_1-\omega_2$ and the latter is equal to $h(t,x)$.
Consequently, we have proved that without loss of generality we
may choose $\omega_{1y}=0$.

Given the condition $\omega_{1y}=0$, equation (\ref{2.3}) reduces
to the relations $A_{2\lambda_i}=0,\ i=1,2$, hence we get $A_2=A_2(\omega_2)$.
Choosing appropriately the function $h_2$ in (\ref{0.6b}) we can put $A_2=0$.
Next, differentiating (\ref{2.2}) with respect to $\lambda_i$ we arrive at
the equations
\begin{equation}
\label{2.8}
A_{0\lambda_i}+A_{1\lambda_i}(\omega_{1t} + y\omega_{1x})=\nu A_{3\lambda_i}
\omega_{2y}^2,\quad i=1,2.
\end{equation}
Differentiating twice the above equations with respect to $y$ yields
\begin{equation}
\label{2.9}
A_{3\omega_2\omega_2\lambda_i}\omega_{2y}^4 +
5A_{3\omega_2\lambda_i}\omega_{2y}^2\omega_{2yy}+
2A_{3\lambda_i}(\omega_{2yy}^2+\omega_{2y}\omega_{2yyy})=0,
\end{equation}
where $i=1,2$.

Note that due to (\ref{2.8}) the inequality $A_{3\lambda_i}\ne 0$ holds.
Dividing (\ref{2.9}) into $A_{3\lambda_i}$ and differentiating the
equality obtained by $\lambda_j, j=1,2$ we get
\begin{equation}
\label{2.10}
\left({A_{3\omega_2\omega_2\lambda_i}\over A_{3\lambda_i}}\right)_{\lambda_j}
\omega_{2y}^2 + 5\left({A_{3\omega_2\lambda_i}\over A_{3\lambda_i}}\right)_{\lambda_j}
\omega_{2yy}=0,\quad i,j=1,2.
\end{equation}
{\bf Case 1.}\ At least one of the four expressions
\[
\left({A_{3\omega_2\lambda_i}\over A_{3\lambda_i}}\right)_{\lambda_j}
\]
does not vanish. Then it is easy to become convinced that
the relation
\[
{\omega_{2yy}\over \omega_{2y}^2}=f(\omega_2)
\]
holds true. Integration of this relation yields
\[
\omega_{2y}=g_1(t,x)\exp\left\{\int f(\omega_2) {\rm d}\omega_2\right\},
\]
where $g_1(t,x)$ is an arbitrary smooth function.

Next, by using the equivalence relation (\ref{0.6a}) we reduce
the equation obtained to the form
\[
\omega_{2y}=g(t,x),
\]
hence
\begin{equation}
\label{2.11}
\omega_2=yg_1(t,x)+g_2(t,x),
\end{equation}
$g_2(t,x)$ being an arbitrary smooth function.

In view of this result, (\ref{2.9}) takes the form
$A_{3\omega_2\omega_2\lambda_i}=0, i=1,2$, hence
\begin{equation}
\label{2.12}
A_3=\Lambda_1(\lambda_1,\lambda_2)\omega_2 + \Lambda_2(\lambda_1,\lambda_2) +
F(\omega_2),
\end{equation}
where $\Lambda_1, \Lambda_2, F$ are arbitrary smooth functions of the
indicated variables. Furthermore, it is not difficult to prove that $\Lambda_1,
\Lambda_2$ are functionally independent (since otherwise the condition (\ref{rank})
would be broken) and, consequently, after redefining $\lambda_1, \lambda_2$
we can represent (\ref{2.12}) in the form
\begin{equation}
\label{2.12a}
A_3=\lambda_1\omega_2 +\lambda_2 + F(\omega_2).
\end{equation}
{\bf Case 2.}\ Suppose that now
\[
\left({A_{3\omega_2\lambda_i}\over A_{3\lambda_i}}\right)_{\lambda_j}=0,\quad
i,j=1,2.
\]
Integrating the above system of PDEs gives the
following form of $A_{3\lambda_i}$
\begin{equation}
\label{2.13}
A_{3\lambda_i}=B_i(\omega_2)L_i(\lambda_1,\lambda_2),\quad i=1,2,
\end{equation}
where $B_i, L_i$ are arbitrary smooth functions and, what is more,
$B_1^2+B_2^2\ne 0$.

As a compatibility condition of system (\ref{2.13}) we get
\[
B_1L_{1\lambda_2}=B_2L_{2\lambda_1}.
\]
{\bf Subcase 2.1.}\  $L_{1\lambda_2}\ne 0,\quad L_{2\lambda_1}\ne 0$.
Given this restrictions the compatibility condition is transformed to
\begin{equation}
\label{2.14}
{B_1(\omega_2)\over B_2(\omega_2)}= {L_{2\lambda_1}\over L_{1\lambda_2}}=
{\rm const}.
\end{equation}
Integrating system (\ref{2.13}) with account of (\ref{2.14}) yields
\[
A_3=\Lambda(\lambda_1,\lambda_2)F_1(\omega_2) + F_2(\omega_2),
\]
where $\Lambda, F$ are arbitrary smooth functions of the indicated variables.
After redefining separation parameters $\lambda_1, \lambda_2$ we
represent the relation as follows
\begin{equation}
\label{2.15}
A_3=\lambda_1 F_1(\omega_2) + F_2(\omega_2).
\end{equation}
{\bf Subcase 2.2}\ $L_{1\lambda_2} = 0,\quad L_{2\lambda_1} = 0$. Integrating
system (\ref{2.13}) and redefining the separation parameters $\lambda_1,
\lambda_2$ yield
\begin{equation}
\label{2.16}
A_3=\lambda_1 S_1(\omega_2) + \lambda_2S_2(\omega_2) + S_0(\omega_2),
\end{equation}
where $S_1, S_2, S_0$ are arbitrary smooth functions. An analysis of
formulae (\ref{2.12a}), (\ref{2.15}) and (\ref{2.16}) shows that the first
two of them are particular cases of formula (\ref{2.16}). Thus, the most
general form of the function $A_3$ is given by (\ref{2.16}).

Inserting (\ref{2.16}) into (\ref{2.8}) and differentiating the
equality obtained with respect to $x$ and $\lambda_j$ gives
$A_{1\lambda_i\lambda_j}=0, i,j=1,2$. Hence, we get for $A_1$
\begin{equation}
\label{2.17}
A_1=\lambda_1 L_1(\omega_1) + \lambda_2 L_2(\omega_1) + L_0(\omega_1),
\end{equation}
where $L_1, L_2, L_0$ are arbitrary smooth functions.

Next, inserting (\ref{2.16}), (\ref{2.17}) into (\ref{2.8}) and differentiating the
equation obtained with respect to $\lambda_j$ we get $A_{0\lambda_i\lambda_j}=0,
i,j=1,2$, hence
\begin{equation}
\label{2.18}
A_0=\lambda_1 R_1(t) + \lambda_2 R_2(t) + R_0(t),
\end{equation}
where $R_1, R_2, R_0$ are arbitrary smooth functions.

With these results we can split equations (\ref{2.1}) and
(\ref{2.2}) by $\lambda_1, \lambda_2$ thus obtaining a system of four
nonlinear PDEs for the three functions $\omega_1, \omega_2, Q$
\begin{eqnarray}
&&Q\omega_{2t} + yQ\omega_{2x} = (\nu y + kx)Q \omega_{2y} +
2\nu Q_y \omega_{2y} + \nu Q\omega_{2yy},\label{2.19}\\
&&Q_{t} + QR_0 +QL_0(\omega_{1t} +y\omega_{1x}) + yQ_x =
\nu Q + (\nu y + kx)Q_y \nonumber\\
&&\quad+\nu Q_{yy} + \nu Q S_0 \omega_{2y}^2,\label{2.20}\\
&&R_1 + L_1(\omega_{1t} + y \omega_{1x}) =\nu S_1 \omega_{2y}^2,\label{2.21}\\
&&R_2 + L_2(\omega_{1t} + y \omega_{1x}) =\nu S_2 \omega_{2y}^2.\label{2.22}
\end{eqnarray}
Making an equivalence transformation (\ref{0.6b}) with appropriately chosen
functions we can put $L_0=0, R_0=0$. Next, due to the requirement (\ref{rank})
$S_1S_2\ne 0$.

There are two inequivalent cases $L_2=0$ and $L_2\ne 0$. Since they are
handled in a similar way, we consider in detail the case $L_2=0$ only.
In view of (\ref{rank}) $L_1$ does not vanish. Choosing appropriately the
functions $f_1, f_2$ in (\ref{0.6a}) we can put $L_1=1, S_2=\pm 1$ in formulae
(\ref{2.19})--(\ref{2.22}). Integrating (\ref{2.22}) with account of (\ref{2.21})
yields for $\omega_2$
\begin{equation}
\label{2.23}
\omega_2=R(t)y+F(t,x),\quad R(t)\ne 0,
\end{equation}
where $R, F$ are arbitrary smooth functions and $R_2=\pm \nu R^2$.

Differentiating (\ref{2.21}) twice with respect to $y$ and taking into account
(\ref{2.23}) we arrive at the equation $S_{1\omega_2\omega_2}=0$, therefore
\[
S_1=C_1\omega_2 + C_2,
\]
where $C_1\ne 0, C_2$ are arbitrary constants. Next, integrating (\ref{2.21}) we
obtain for $\omega_1, F(t,x)$
\begin{equation}
\label{2.24}
\begin{array}{l}
\omega_1=\nu C_1 (R^3 x + P(t))- \int R_1(t) {\rm d}t,\\[2mm]
F(t,x)=3\dot R +R^{-2}\dot P(t) -C_1^{-1} C_2,
\end{array}
\end{equation}
where $P(t)$ is an arbitrary smooth function.

Hence we conclude that the corresponding solution with separated variables
reads as
\begin{eqnarray*}
u&=&Q(t,x,y)\exp\{\lambda_1\int R_1(t) {\rm d}t + \lambda_2\int R_2(t) {\rm d}t\}
\exp\{\lambda_1 \omega_1\} \varphi_2(\omega_2)\\
&=&Q(t,x,y)\exp\{\lambda_2\int R_2(t) {\rm d}t\} \exp\{\lambda_1 (\nu C_1 (R^3 x +
P(t)))\}\varphi_2(\omega_2).
\end{eqnarray*}
Thus, the function $R_1(t)$ does not enter into the solution with separated
variables and, therefore, we can put $R_1=0$ in (\ref{2.24}). Furthermore,
within an equivalence transformation (\ref{0.6a}) we can choose
$C_1=\nu^{-1}, C_2=0$, thus getting
\begin{eqnarray}
\omega_1&=&R(t)^3 x + P(t),\label{2.25}\\
\omega_2&=&R(t)y + 3 \dot R(t) x + \dot P R(t)^{-2}.\label{2.26}
\end{eqnarray}

Provided $L_2\ne 0$, the forms of the functions $\omega_1, \omega_2$ is
the same as those given in (\ref{2.25}), (\ref{2.26}).

Inserting (\ref{2.25}) and (\ref{2.26}) into (\ref{2.19}) and integrating
by $y$ we get the form of the factor $Q(t,x,y)$
\begin{equation}
\label{2.27}
\begin{array}{rcl}
Q&=&\exp\Biggl\{\dis \left({4\dot R - \nu R\over 4\nu R}\right)y^2 +
{3\ddot R -kR\over 2\nu R}xy +
{y\over 2\nu R} {d \over dt}\left({\dot P\over R^2}\right)\\[3mm]
&&+ M(t,x)\Biggr\}.
\end{array}
\end{equation}
Substituting (\ref{2.27}) into (\ref{2.20}) we come to the following
relation
\begin{eqnarray}
&&
{1\over\nu} {d\over dt}\left( {\dot R\over R}\right)y^2+{3\over 2\nu}
 {d\over dt}\left( {\ddot R\over R}\right)xy+{1\over 2\nu}\dot Zy+M_t+
{1\over 2\nu}\left({3{\ddot R\over R}-k} \right)y^2+yM_x\nonumber\\
&&\
={\nu\over 2}+2{\dot R\over R}+(\nu y+kx)\left({\left({{2\dot R \over \nu
R}-{1\over 2}}\right)y+{1\over 2\nu}\left({3{\ddot R\over R}-k} \right)x+
{1\over 2\nu}Z}\right)\nonumber\\
&&\quad
+\nu\left({\left({{2\dot R\over \nu R}-{1\over 2}}\right)y+{1\over 2\nu}
\left({3{\ddot R\over R}-k} \right)x+{1\over 2\nu}Z}\right)^2+\nu S_0R^2,
\label{2.28}
\end{eqnarray}
where we use the notation
\[
Z(t)=R^{-1}{{\rm d}\over {\rm d}t}\left({\dot P\over R^2}\right).
\]

Differentiating (\ref{2.28}) three times with respect to $y$ yields
$S_{0\omega_2\omega_2\omega_2}=0$, therefore
\[
S_0=C_1 \omega_2^2 + C_2\omega_2 + C_3,
\]
where $C_1, C_2, C_3$ are arbitrary constants. Next, differentiating
(\ref{2.28}) with respect to $y$ twice and with respect to $x$ once
we get $M_{xxx}=0$, or
\[
M=M_1(t)x^2+M_2(t)x + M_3(t),
\]
where $M_1, M_2, M_3$ are arbitrary smooth functions.

Finally, inserting the obtained expressions for $S_0, M$ into (\ref{2.28}) and
splitting by the variables $x, y$ we come to the following system of ODEs:
\begin{eqnarray}
&&{\ddot R\over R} =2{\dot R^2\over R^2}+{2\nu^2\over 5}C_1R^4-{\nu^2\over 10}
+{k\over 5}, \label{2.29}\\
&&M_1=-{3\stackrel{...}{R}\over 4\nu R}+{15\over 4\nu}+3\nu C_1\dot RR^3-
{k\over 4},  \label{2.30}\\
&&\dot M_1={9\ddot R^2\over 4\nu R^2}+9\nu C_1R^2\dot R^2-{k^2\over 4\nu},
 \label{2.31}\\
&&M_2=-{1\over 2\nu}\dot Z+{2\dot R\over \nu R}Z+\nu C_2R^3+2\nu R\dot PC_1,
\label{2.32}\\
&&\dot M_2={3\ddot R\over 2\nu R}Z+3\nu R^2\dot RC_2+6\nu \dot R\dot PC_1,
\label{2.33}\\
&&\dot M_3={\nu\over 2}+2{\dot R\over R}+{1\over  4\nu}Z^2+\nu C_1{\dot P^2
\over \dot R^2}+\nu C_2\dot P+\nu  C_3R^2.
\label{2.34}
\end{eqnarray}

Differentiating (\ref{2.30}) with respect to $t$ and subtracting
the resulting equation from (\ref{2.31}) yields the fourth-order ODE
for the function $R$
\[
-{R^{(IV)}\over R} + 6 {\dot R\stackrel{...}{R}\over R^3} + 2{\ddot R^2\over R^2}
-10{\dot R^2\ddot R\over R^3}+ 4\nu^2 C_1\ddot R R^3 + \frac{k^3}{3}=0.
\]
Reducing the order of the above ODE with the help of equation (\ref{2.29})
and its first- and second-order differential consequences we arrive at the
following relation:
\begin{equation}
\label{2.35}
{4\nu^2\over 25}C_1^2 R^8 = {\nu^4\over 100} + {k^2\over 25} - {\nu^2 k\over 25}-
{k^2\over 9}.
\end{equation}

If in (\ref{2.35}) $C_1 \ne 0$, then in view of (\ref{2.29}) $k=0$. Provided,
$C_1=0$, $k$ is a root of the quadratic equation
\[
64k^2 + 36\nu^2k - 9\nu^2=0,
\]
hence $k=3\nu^2/16$ or $k=-3\nu^2/4$.

Thus system of ODEs (\ref{2.29})--(\ref{2.34}) is consistent only if
the parameter $k$ takes one of three values $0,\ 3\nu^2/16,\
-3\nu^2/4$.  Consequently, KE (\ref{0.1}) has solutions with separated
variables in the case considered (i.e., provided the system
(\ref{0.3}) takes the form (\ref{0.4})) only for the values of the
parameter $k$ given previously. This provides the proof of the first part
of Theorem 1.

We examine the three possible cases $0,\ 3\nu^2/16,\ -3\nu^2/4$ separately.
\vspace{2mm}

\noindent
{\bf Case 1.}\ For $k=0$. Then the equality $R(t)=\pm 2^{-1/2}S_1^{-1/4}={\rm const}$
holds. We denote this constant as $r$. Next, it follows from (\ref{2.33}) that
$M_2=m={\rm constant}$. In view of these facts we get from (\ref{2.32}) ODE for $P(t)$
\[
-\stackrel{...}{P} + \nu^2\dot P + 2\nu r^3 (\nu S_2 r^3 - m) =0
\]
which general solutions reads
\begin{equation}
\label{2.36}
P(t)=C_4{\rm e}^{\nu t} + C_5{\rm e}^{-\nu t} + 2r^3 (m\nu^{-1} - S_2 r^3)t + C_6,
\end{equation}
where $C_4, C_5, C_6$ are arbitrary constants.

A direct check shows that applying finite transformations from the
symmetry group admitted by KE under $k=0$ to the obtained solution
with separated variables (\ref{0.2}), (\ref{2.25}), (\ref{2.26}),
(\ref{2.36})  we can cancel $P(t)$.

Scaling when necessary $\omega_1, \omega_2$ in (\ref{2.25}), (\ref{2.26}) we
can choose $r=1$. Hence we get the equality $C_1={1\over 4}$. Summing up we
conclude that the following relations hold:
\begin{eqnarray*}
&&Q=\exp\left({-{y^2\over 4}+\nu C_2x+\nu\left({C_3+{1\over 2}}\right)t}\right),\quad
\omega_1=x, \quad \omega_2=y;\\
&&\dot \varphi_0 =\nu\left({\lambda_1-\left({C_3+{1\over 2}}\right)}\right)
\varphi_0,\\
&&\dot \varphi_1 =\nu(\lambda_2-C_2)\varphi_1,\quad
\ddot \varphi_2 =\left({{\omega_2^2\over 4}+\lambda_2\omega_2+\lambda_1-{1
\over 2}}\right)\varphi_2.
\end{eqnarray*}
Then the corresponding solution with separated variables is
$$
u=\varphi_2\exp\left({-{y^2\over 4}+\nu (\lambda_1t+\lambda_2x)}\right).
$$
Consequently, the constants $C_2$ and $C_3+{1\over 2}$ do not enter the
final form of the solution with separated variables. This means that we
can put $C_2=0$ and $C_3=-{1\over 2}$.

Thus we have proved the validity of the first part of Theorem 5.
\vspace{2mm}

\noindent
{\bf Cases 2,3.}\ For $k=\frac{3\nu^2}{16}$ or $k=-\frac{3\nu^2}{4}$. In
these cases we get from (\ref{2.29})
\[
{\ddot R\over R} -2\left({\dot R\over R}\right)^2=-a^2,
\]
where
\[
a=\left\{\begin{array}{ll} {\dis \frac{\nu}{4}},& {\rm under}\
k={\dis \frac{3\nu^2}{16}},\\[3mm]
{\dis \frac{\nu}{2}},& {\rm under}\
k=-{\dis \frac{3\nu^2}{4}}.\end{array}\right.
\]
Integrating the above ODEs yields
\[
R(t)=(C_1\sinh at + C_2\cosh at)^{-1},
\]
where $C_1, C_2$ are arbitrary constants.

Using shifts with respect to $t$ and the equivalence transformation
(\ref{0.6a}) we get the four inequivalent forms of the function $R(t)$
\[
R(t)={1\over \cosh at},\quad R(t)={1\over \sinh at},\quad
R(t)=\exp\{\pm at\}.
\]

Comparing (\ref{2.32}) and the first-order differential consequence of
(\ref{2.33}) yields the second-order ODE for $Z(t)=R^{-1} ({\rm d}/{\rm d}t)
(\dot P/R^2)$
\begin{equation}
\label{2.37}
-\ddot Z + 4{\dot R\over R}\dot Z + \left({\ddot R\over R} - 4
{\dot R\over R}\right)Z = 0.
\end{equation}

The general solution of this equation has the following structure:
\[
Z(t)=C_1 Z_1(t) + C_2 Z_2(t),
\]
where $C_1, C_2$ are integration constants. Hence, we conclude
that the function $P(t)$ is of the form
\begin{equation}
\label{2.38}
P(t)=C_1 P_1(t) + C_2 P_2(t) + C_3 P_3(t) + C_4 P_4(t),
\end{equation}
where $C_3, C_4$ are integration constants.

On the other hand, if we apply to the solution with separated variables
(\ref{0.2}), (\ref{2.25}), (\ref{2.26}) with $P(t)=0$ finite transformations
from the symmetry group of KE under $k=\frac{3\nu^2}{16}$ or $k=-\frac{3\nu^2}{4}$,
then we get an equivalent solution with separated variables such that
$P(t)$ is of the form
\begin{equation}
\label{2.39}
P(t)=C_1' P_1'(t) + C_2' P_2'(t) + C_3' P_3'(t) + C_4' P_4'(t).
\end{equation}
Here $C_1',\ldots, C_4'$ are arbitrary constants and the functions
$P_1'(t),\ldots, P_4'(t)$ are linearly independent. Hence, we conclude
that due to the theorem on existence and uniqueness of the Cauchy
problem for a fourth-order ODE (\ref{2.37}) (considered as an
equation for the function $P(t)$) the expressions on the right-hand
sides of (\ref{2.38}) and (\ref{2.39}) coincides within the choice
of constants $C_i, C_i', i=1,\ldots, 4$. Consequently, without
loss of generality we can put $P(t)=0$ in formulae (\ref{2.25}),
(\ref{2.26}).

Using the reasonings analogous to those of Case 1 we can put $r=1, C_2=0, C_3=0$.
The second part of Theorem 6 is proved.

A similar analysis of the separability of KE into three ODEs (\ref{0.5})
yields the proofs of the remaining assertions from Section 2.

\section{Exact solutions}

Remarkably, for the equation under study it is possible to give a complete
account of solutions with separated variables. For the case when KE separates
into three first-order ODEs (\ref{0.5}), we get the following family of its exact
solutions:
\begin{eqnarray*}
u&=&\exp\Biggl\{ \nu\int\left({f_1\lambda_1+f_2\lambda_2\over  \dot f_2f_1-\dot f_1
f_2}\right)^2{\rm d}t+\lambda_1{f_1y-\dot f_1x\over \dot f_2f_1-\dot f_1f_2}+
\lambda_2{f_2y-\dot f_2x\over \dot f_2f_1-\dot f_1f_2}+\\
&&+\left({-{1\over 4\nu}{\ddot f_2f_1-\ddot f_1f_2\over \dot f_2
f_1-\dot f_1f_2}-{1\over 4}}\right)y^2+{1\over 2\nu} \left({
{\ddot f_2\dot f_1-\ddot f_1\dot f_2\over \dot f_2f_1-\dot f_1f_2}-k}\right)
xy+\\
&&+\left({{1\over 4\nu}{\stackrel{...}{f_2}\dot f_1-\stackrel{...}{f_1}\dot
f_2\over \dot f_2f_1-\dot f_1f_2}-{k\over 4}}\right)x^2-{1\over 2}\ln|
 \dot f_2f_1-\dot f_1f_2|+{\nu\over 2}t\Biggl\},\\
\end{eqnarray*}
where $k, f_1(t), f_2(t)$ are given by the corresponding formulae from
Theorems 2--5.

Next, for the case when KE separates into three ODEs of the form (\ref{0.4})
we obtain the following families of its exact solutions:
\begin{enumerate}
\item {$k=0$ (this case has been considered in Theorem 5)
\[
u=\exp\left({-{y^2\over 4}+\nu(\lambda_1t+\lambda_2x)}\right)\,
D_{\lambda_2^2-\lambda_1}(y+2\lambda_2),
\]
where $D_\nu$ is the parabolic cylinder function.}
\item{$k=3\nu^2/16$ or $k=-3\nu^2/4$
\begin{eqnarray*}
u&=&\exp\Biggl\{\nu\lambda_1\int R^2{\rm d}t+\nu\lambda_2R^3x\left({{\dot R\over \nu
 R}-{1\over 4}}\right)y^2+{1\over 2\nu}\left({3{\ddot R\over R}-k}\right)xy\\
&&+\left({-{3\stackrel{...}{R}\over 4\nu R}+{15\dot R\ddot R\over 4\nu R^2}-
{k\over 4}}\right)x^2+{\nu\over 2}t+2\ln R\Biggr\}\{\lambda_2(Ry+3\dot Rx)\\
&&+\lambda_1\}^{\frac{1}{2}}
Z_{1\over 3}\left({{2\over 3\lambda_2}(\lambda_2(Ry+3\dot Rx)+\lambda_1)^
{3\over 2}}\right),
\end{eqnarray*}}
\end{enumerate}
where $R$ is given by the corresponding formula from Theorem 6 and
$Z_{1\over 3}$ is the cylindric function.

Note that the above obtained families of exact solutions of KE
contain two continuous parameters $\lambda_1, \lambda_2$. These
parameters have the meaning of eigenvalues of two
commuting symmetry operators of KE, while the corresponding
solution with separated variables is the eigenfunction of
these operators. Provided some appropriate boundary and initial
conditions are imposed, the parameters become discrete and
thus we get a basis for expanding sufficiently smooth
solutions of KE into series.

\section{Conclusions}

It is a remarkable feature of the Kramers equation (\ref{0.1}) that
a classical problem of variable separation can be solved in full
generality. The results obtained on this way are in good
correspondence with the ones on symmetry classification of KEs
of the form (\ref{0.1}). As follows from the papers
\cite{spi97,spi98}, the cases $k=3\nu^2/16, k=-3\nu^2/4$ are
distinguished by the fact that the corresponding KEs (\ref{0.1})
admit the most extensive symmetry groups. For these choices
of $k$ KE (\ref{0.1}) is invariant with respect to eight-parameter
Lie transformation groups, while for all other values of the
parameter $k$ the maximal group is six-parameter.

\section*{Acknowledgments} One of the authors (R.Zh.) is partially supported
by the Alexander von Humboldt-Stiftung and Technical University of
Clausthal.

\end{document}